\documentclass[superscriptaddress,twocolumn,prd,aps,showpacs,secnumarabic]{revtex4}
\usepackage{graphicx}
\begin{document}

\title{A note on the rare decay of a Higgs boson into photons and a $Z$ boson}
\author{Ali Abbasabadi}
\affiliation{Department of Physical Sciences,
Ferris State University, Big Rapids, Michigan 49307, USA}
\author{Wayne W. Repko}
\affiliation{Department of Physics and Astronomy,
Michigan State University, East Lansing, Michigan 48824, USA}

\date{\today}
\begin{abstract}
\vspace*{0.5cm} \hspace*{0.1cm} We have calculated the width of the rare decay $H\to\gamma\gamma Z$ at one-loop level in the standard model for Higgs boson masses in the range $115\, {\rm GeV} \leq m_H \leq 160\, {\rm GeV}$ . For this range of Higgs boson masses we find that $Z$ boson is predominantly  longitudinally polarized, and the photons have the same helicity. A comparison of the decay width $\Gamma(H\rightarrow \gamma\gamma Z)$ to those of $H\to\gamma\gamma$ and $H\to\gamma Z$ shows that, for the Higgs boson mass of $m_H \sim 135\, \rm{GeV}$, the ratios of the decay widths are $\Gamma(H\to\gamma\gamma Z) / \Gamma(H\to\gamma\gamma) \sim \Gamma(H\to\gamma\gamma Z) / \Gamma(H\to\gamma Z) \sim 10^{-5}-10^{-6}$.
\end{abstract}
\maketitle

\vskip1pc

\section{Introduction}
\label{sec:1} 
Due to the lack of a direct coupling between the $Z$ boson and
the photon in the standard model, the lowest order contribution to the decay $H\to\gamma\gamma Z$ occurs at the one-loop level. The Feynman diagrams for this process are identical to those of the crossed channel scattering process $\gamma\gamma\to Z H$, which was studied by Gounaris, Porfyriadis, and Renard \cite{gpr}. The decay $H\to\gamma\gamma$, which also takes place at the one-loop level and is usually viewed as a discovery mode for an intermediate mass Higgs boson \cite{ghkd}, dominates the decay process $H\to\gamma\gamma Z$ by several orders of magnitude. However, by imposing kinematic cuts on the photons and $Z$ boson in the decay $H\to\gamma\gamma Z$, we may exclude contributions of the back-to-back photons. It is therefore in principle possible to distinguish the photons from $H\to\gamma\gamma$ decay from those arising from the decay $H\to\gamma\gamma Z$. While admittedly very rare, the decay $H\to\gamma\gamma Z$, among one loop decay processes, is particularly sensitive to top quark couplings.

In the next section, we discuss the calculations of the decay width, the photon invariant mass decay distribution, and the decay energy distribution of the $Z$. This is followed by a summary and conclusions.

\section{Decay Width Calculation}
\label{sec:2}

\begin{figure}[h]
\centering\includegraphics[width=2.5in]{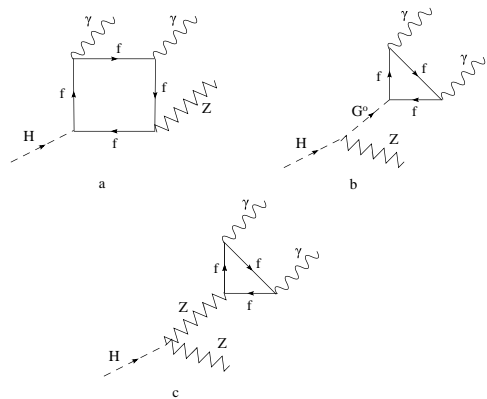}
\caption{\footnotesize The generic Feynman diagrams for the decay process $H\to\gamma\gamma Z$ are shown. For the charged fermion $f$ in the Figs.\,\ref{feyndiag}(a) and \ref{feyndiag}(b), we include only the top quark. However, in the Fig.\,\ref{feyndiag}(c), the $Z$ boson exchange, all charged fermions $f$ are included in the loop.} \label{feyndiag}
\end{figure}

To facilitate our calculations of the amplitudes for the process $H\to\gamma\gamma Z$, we use a generalized non-linear gauge fixing condition \cite{bc}. In the gauge fixing terms there are several parameters which can be set freely without affecting the values of the measurable quantities. We set the values of these
parameters in such a way as to minimize the number of Feynman diagrams contributing to the $H\to\gamma\gamma Z$ amplitudes. In addition, we use the 't Hooft-Feynman gauge, which reduces the propagators for the gauge bosons to a simple form $-ig_{\mu\nu}/(k^2 - m^2)$, where $k$ and $m$ are the momentum and the mass of a gauge boson, respectively. The resulting Feynman diagrams are identical to those of the crossed channel scattering process $\gamma\gamma\to Z H$, which was investigated in the Ref. \cite{gpr} for the standard model and its minimal supersymmetric extension. The generic diagrams 
are drawn \cite{jaxo} in the Fig.\,\ref{feyndiag}.  
As noted in the Ref. \cite{gpr}, due to the charge conjugation properties of the gauge boson couplings there are no $W$ bosons in the loops and we need only consider contributions from charged fermion loops. In the box diagram, Fig.\,\ref{feyndiag}(a), the coupling of the Higgs boson to the fermion in the loop is proportional to the fermion mass and we include only the top quark contribution. The same thing is true in the triangle diagram with the neutral Goldstone boson $G^0$ exchange, Fig.\,\ref{feyndiag}(b), and, again we include only the top quark contribution. 

In the $Z$ boson exchange diagram, Fig.\,\ref{feyndiag}(c), we encounter an anomalous contribution \cite{a}, whose cancellation requires the inclusion of all charged fermions of a given generation. After the cancellation of the anomalous contribution, which is independent of the $m_f$, the mass of the charged fermion in the loop, the contribution of this diagram to the dominant helicity amplitudes is proportional to the $m^2_f$. As mentioned above, the contributions of Fig.\,\ref{feyndiag}(a) and Fig.\,\ref{feyndiag}(b) to any helicity amplitude is proportional to the $m^2_t$ \cite{gpr}. We, therefore, could have included only the charged fermions of the third generation in the evaluation of diagram Fig.\,\ref{feyndiag}(c) but chose to include all three generations in this diagram. 

For the decay $H\to\gamma\gamma Z$, the helicity amplitudes ${{\cal A}_{\lambda_{1}\lambda_{2}\lambda_{Z}}}$, where $\lambda_{1}$ and $\lambda_{2}$ are the helicities of the photons and $\lambda_{Z}$ is the helicity of the $Z$ boson, are not independent. Due to Bose symmetry and $CP$ invariance, there are only four independent helicity amplitudes \cite{gpr}, which we choose to be ${\cal A}_{++0}$, ${\cal A}_{+++}$, ${\cal
A}_{+--}$, and ${\cal A}_{+-0}$. Any other helicity amplitude is expressible in terms of these four amplitudes. For instance, $CP$ invariance relates ${\cal A}_{--0}$ to the basic set as
\begin{equation}
{\cal A}_{--0}  =  - {\cal A}_{++0}\,.\label{cp}\
\end{equation}
Our numerical calculation of the amplitudes indicates that, among the four basis amplitudes, the helicity amplitude ${\cal A}_{++0}$ is by far the largest. As a consequence, we need only consider the amplitudes ${\cal A}_{++0}$ and ${\cal A}_{--0}$.

Our calculations were performed in a {\it semiautomatic} method \cite{hs}. Specifically, we used the FeynArts package \cite{h}, to generate the helicity amplitudes in terms of the tensor loop integrals. Using the algebraic manipulation program {\sc{form}} \cite{form}, these tensor integrals were converted to the conventions of Passarino-Veltman \cite{pv} and expressed in terms of scalar integrals. The scalar integrals were calculated using code written following the approach of 't Hooft-Veltman \cite{tv} and Denner-Nierste-Scharf \cite{dns}.  As a partial test of our calculations, we compared the numerical results of our routines for the scalar integrals with those of the {\sc{fortran}} codes {\sc{loop}} \cite{kd} and the {\sc{ff}} \cite{ff}. While these two packages did not give numerically stable results for all regions of the parameter space, where comparisons were possible we obtained essentially  identical numerical results for our scalar integrals and those of the {\sc{loop}} and/or {\sc{ff}}. We also checked the gauge invariance of the helicity amplitudes for each of the final state photons.

To facilitate the discrimination of $H\to\gamma\gamma Z$ from $H\to\gamma\gamma$ and $H\to\gamma Z$ and account for possible experimental limitations, we imposed a variety of cuts on the following photon and $Z$ boson variables: $|\vec{p}_\gamma|$, $|\vec{p}_{\gamma'}|$, $|\vec{p}_Z|$, $m_{\gamma\gamma'}^2$, $m_{\gamma  Z}^2$, $m_{\gamma' Z}^2$, $\theta_{\gamma\gamma'}$, $\theta_{\gamma Z}$, and $\theta_{\gamma' Z}$. Here, $\vec{p}_\gamma$, $\vec{p}_{\gamma'}$, and $\vec{p}_{Z}$ are the 3-momenta of the photons and the $Z$ boson, respectively, in the center of mass of the Higgs boson and $\theta_{\gamma\gamma'}$, $\theta_{\gamma Z}$, and $\theta_{\gamma' Z}$ are the various angles between the 3-momenta, $\vec{p}_\gamma$, $\vec{p}_{\gamma'}$, and $\vec{p}_Z$. The invariant mass variables are $m_{\gamma\gamma'}^2 = (p_{\gamma}+p_{\gamma'})^2$, $m_{\gamma Z}^2 = (p_{\gamma}+p_{Z})^2$, and $m_{\gamma' Z}^2 = (p_{\gamma'}+p_{Z})^2$. Here, $p_{\gamma}$, $p_{\gamma'}$, and $p_{Z}$ are the 4-momenta of the photons and the $Z$ boson, respectively.

For our calculations of the decay width $\Gamma(H\rightarrow \gamma\gamma Z)$, invariant mass distribution $d\Gamma(H\rightarrow \gamma\gamma Z)/dm_{\gamma\gamma'}$, and the $Z$ boson energy distribution $d\Gamma(H\rightarrow \gamma\gamma Z)/dE_{Z}$, we choose the following set of cuts:
\begin{eqnarray}
|\vec{p}_{\gamma}|_{\rm cut}=
|\vec{p}_{\gamma'}|_{\rm cut}=
|\vec{p}_{Z}|_{\rm cut} &\equiv& |\vec{p}\,|_{\rm cut} \,,\label{pcut}\\
(m_{\gamma\gamma'})_{\rm cut}=
(m_{\gamma Z})_{\rm cut}=
(m_{\gamma' Z})_{\rm cut} &\equiv&  m_{\rm cut}\,,\label{mcut}\\
(\theta_{\gamma\gamma'})_{\rm cut}=
(\theta_{\gamma Z})_{\rm cut}=
(\theta_{\gamma' Z})_{\rm cut} &\equiv& \theta_{\rm cut} \,.\label{tcut}
\end{eqnarray}
These cuts facilitate the experimental tagging of the photons and the $Z$ boson. They provide minimum opening angles between the photons and the $Z$ boson, exclude contributions of the back-to-back photons, and also avoid any numerical instability in the calculations. The cuts help discriminate the non-back-to-back photon pairs of the decay $H\rightarrow  \gamma\gamma Z$ from the back-to-back $\gamma\gamma$ pairs in the decay $H\rightarrow \gamma\gamma$. In principle, all the photons of the decays $H\rightarrow \gamma\gamma$, $H\rightarrow \gamma Z$, and $H\rightarrow \gamma\gamma Z$ can be identified.

\begin{figure}[h]
\centering\includegraphics[width=2.5in]{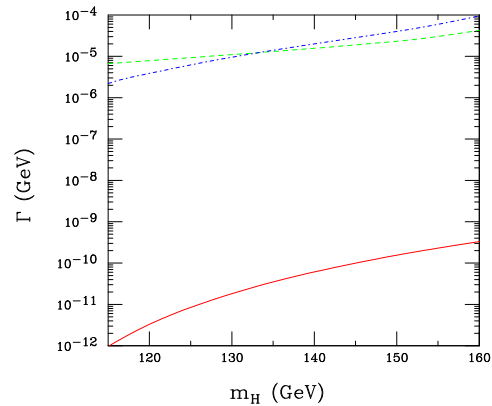}
\caption{\footnotesize (Color online) The decay widths as function of  $m_H$ for three decay modes of the Higgs boson are shown. The solid line is $\Gamma(H\to\gamma\gamma Z)$, the dashed line is  $\Gamma(H\to\gamma Z)$, and the dotdashed line is $\Gamma(H\to\gamma\gamma)$. The cuts imposed on $\Gamma(H\to\gamma\gamma Z)$ are $|\vec{p}\,|_{\rm cut}= 5\, \rm{GeV}$, $m_{\rm cut}= 10\, \rm{GeV}$, and $\theta_{\rm cut} = \frac{\pi}{24}$.} \label{decaywidth}
\end{figure}

In Fig.\,\ref{decaywidth}, we show the result of the calculation for the decay width $\Gamma(H\rightarrow \gamma\gamma Z)$ as a function of the Higgs boson mass $m_H$. 
For purposes of comparison, in this figure we also include the decay widths $\Gamma(H\rightarrow \gamma\gamma)$ and $\Gamma(H\rightarrow \gamma Z)$ \cite{vvss}, which are calculated using the {\sc{hdecay}} package \cite{hdecay,gmr}. As it is clear from this figure, for the Higgs boson masses that we are considering, the decay width $\Gamma(H\rightarrow\gamma\gamma Z)$ is several orders of magnitude smaller than the decay widths $\Gamma(H\rightarrow \gamma\gamma)$ and $\Gamma(H\rightarrow \gamma Z)$. For instance, for a Higgs boson mass of $m_H \sim 135\, \rm{GeV}$, the ratios of the decay widths are $\Gamma(H\to\gamma\gamma Z) / \Gamma(H\to\gamma\gamma) \sim \Gamma(H\to\gamma\gamma Z) / \Gamma(H\to\gamma Z) \sim 10^{-6}$. There are no cuts on the decay widths $\Gamma(H\rightarrow \gamma\gamma)$ and $\Gamma(H\rightarrow \gamma Z)$. The cuts we imposed, of course, decrease the value of $\Gamma(H\rightarrow \gamma\gamma Z)$. In addition, there is the suppression from three-body phase space, and from the higher order in $\alpha$. However, there are other differences in the amplitudes for these decay modes, which account for their relative sizes. For instance, in the case of $H\rightarrow \gamma\gamma$, the decay amplitude receives contributions from charged fermion loops as well as a substantial contribution from $W$  boson loops, whereas in the decay $H\rightarrow \gamma\gamma Z$, there are no $W$  boson loop contributions and  the anomalous triangle diagram, Fig.\,\ref{feyndiag}(c), must be included. As a result of these differences, the simple power counting method for estimating the size of the ratio of the decay widths $\Gamma(H\rightarrow \gamma\gamma Z) / \Gamma(H\rightarrow \gamma\gamma)$ is rather unreliable. 

\begin{figure}[h]
\centering\includegraphics[width=2.5in]{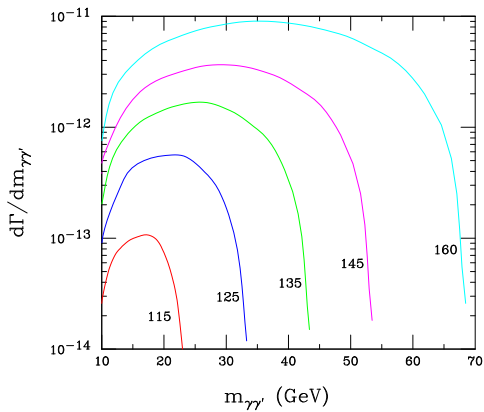}
\caption{\footnotesize (Color online) The invariant mass distributions $\Gamma(H\rightarrow \gamma\gamma Z)/dm_{\gamma\gamma'}$ as function $m_{\gamma\gamma'}$, the invariant mass of the final photons, for Higgs  masses of $m_H$ = 115, 125, 135, 145, and 160 GeV are shown. The cuts imposed are the same as those for the total width $\Gamma(H\rightarrow \gamma\gamma Z)$ of the Fig.\,\ref{decaywidth}.} \label{massdist}
\end{figure}

\begin{figure}[h]
\centering\includegraphics[width=2.5in]{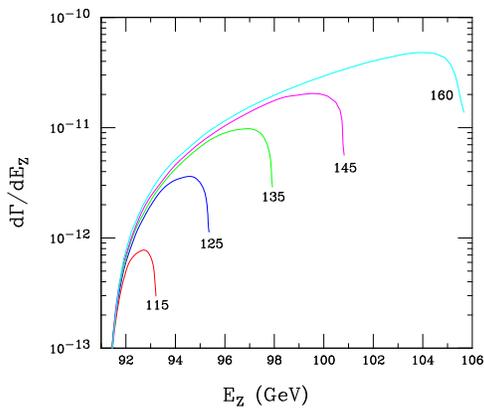}
\caption{\footnotesize (Color online) The energy distributions $\Gamma(H\rightarrow \gamma\gamma Z)/dE_Z$, as function of the $Z$ boson energy $E_Z$, for the Higgs masses of Fig.\,\ref{massdist} are shown. The cuts imposed are the same as those for the total width  $\Gamma(H\rightarrow \gamma\gamma Z)$ of the Fig.\,\ref{decaywidth}.} \label{energydist}
\end{figure}

In Fig.\,\ref{massdist}, we show the invariant mass distribution $d\Gamma(H\rightarrow \gamma\gamma Z)/dm_{\gamma\gamma'}$ as function of $m_{\gamma\gamma'}$ and in Fig.\,\ref{energydist}, we show the energy distribution $d\Gamma(H\rightarrow \gamma\gamma Z)/dE_{Z}$ as function of the $Z$ boson energy $E_Z$.

\begin{figure}[h]
\centering
\begin{minipage}[c]{\textwidth}
\includegraphics[width=2.0in]{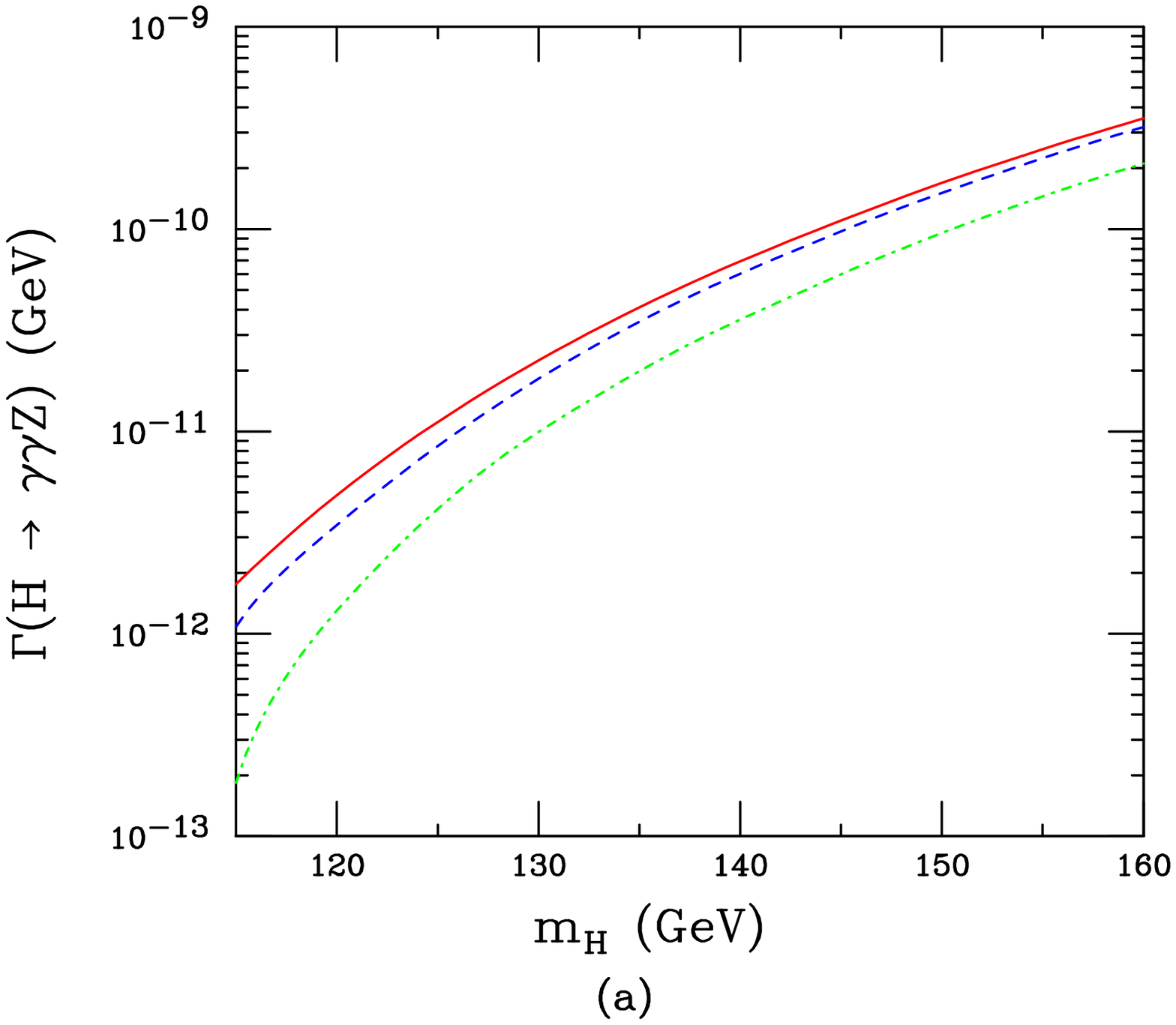}
\hspace{1in}%
\includegraphics[width=2.0in]{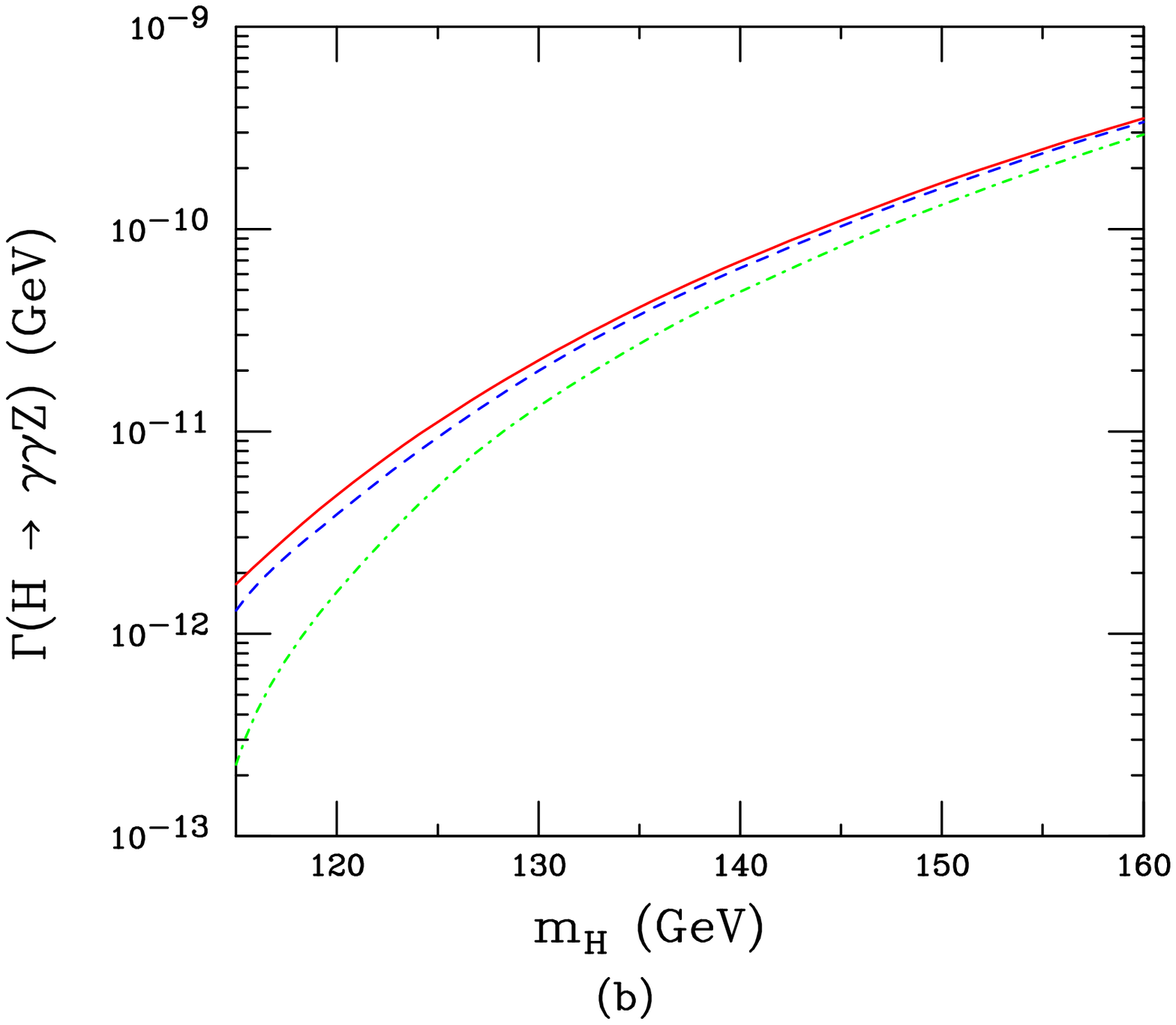}  \\
\includegraphics[width=2.0in]{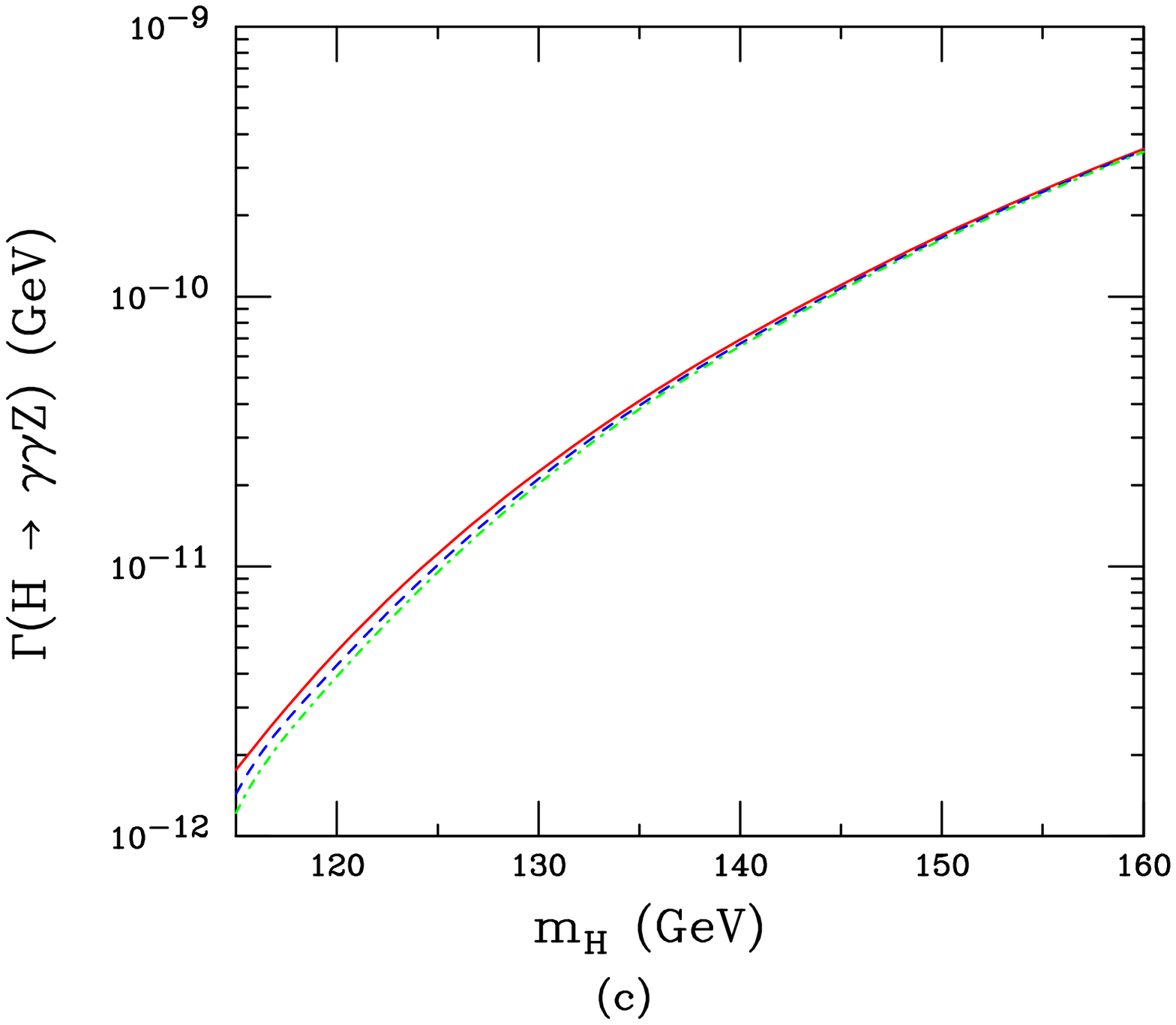}
\hspace{1in}%
\includegraphics[width=2.0in]{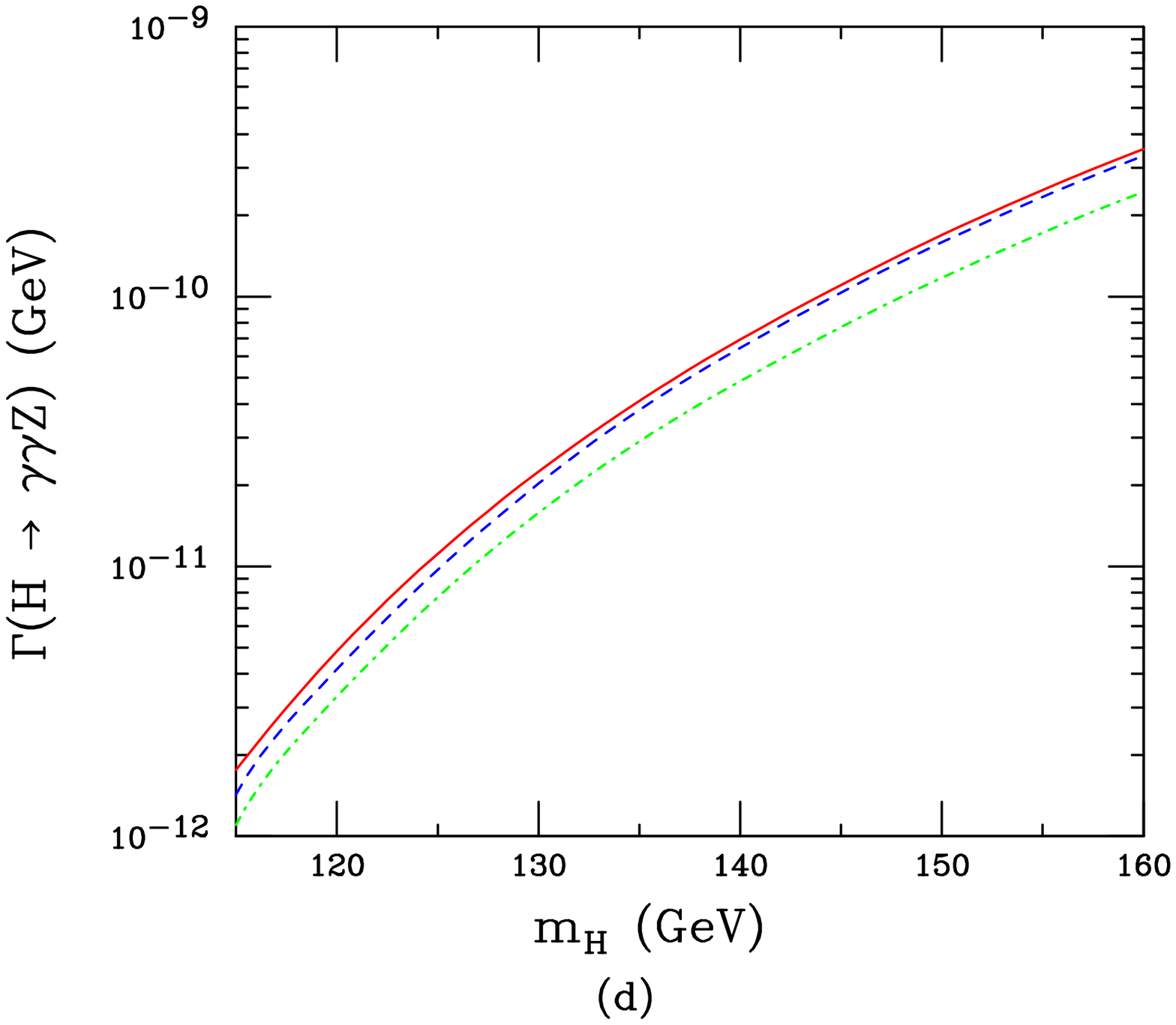}\hfill
\caption{(Color online) Fig.\,\ref{cut}\,(a) shows $\Gamma(H\to\gamma\gamma Z)$ for the cuts ($|\vec{p}\,|_{\rm cut}$, $m_{\rm cut}$,$\theta_{\rm cut}$)=(1 GeV, 1 GeV, $\pi/48$) (solid), (5 GeV, 5 GeV, $\pi/6$) (dashed), (10 GeV, 10 GeV, $\pi/3$) (dotdash). In Fig.\,\ref{cut}\,(b), $m_{\rm cut}$= 1 GeV, $\theta_{\rm cut}=\pi/48$ and $|\vec{p}\,|_{\rm cut}$= 1 GeV (solid), 5 GeV (dashed), 10 GeV (dotdash); in Fig,\.\,\ref{cut}\,(c), $|\vec{p}\,|_{\rm cut}$= 1 GeV, $\theta_{\rm cut}=\pi/48$, and $m_{\rm cut}$= 1 GeV (solid), 5 GeV (blue), 10 GeV (dotdash); and in Fig.\,\ref{cut}\,(d), $|\vec{p}\,|_{\rm cut}$= 1 GeV, $m_{\rm cut}$= 1 GeV, and $\theta_{\rm cut}$=$\pi/48$ (solid), $\pi/6$ (blue ), and $\pi/3$ (dotdash).}\label{cut}
\end{minipage}
\end{figure}

To further investigate the effect of cuts on the decay width $\Gamma(H\rightarrow \gamma\gamma Z)$, in Fig.\,\ref{cut} we plot $\Gamma(H\rightarrow \gamma\gamma Z)$ as function of $m_H$, for several different choices of $|\vec{p}\,|_{\rm cut}$, $m_{\rm cut}$, and $\theta_{\rm cut}$.  In the Fig.\,\ref{cut}\,(a), all three cut parameters $|\vec{p}\,|_{\rm cut}$, $m_{\rm cut}$ and $\theta_{\rm cut}$ are varied simultaneously. This figure captures the combined effect of all three cuts on the decay width.
The effects of the individual cuts on the decay width are illustrated Figs.\,\ref{cut}\,(b)-(d). In Fig.\,\ref{cut}(b), the invariant mass and angular cuts are held fixed and the 3-momentum cuts are allowed to vary, in Fig.\,\ref{cut}\,(c) we hold the 3-momentum and the angular cuts fixed and vary the invariant mass cut, and in Fig.\,\ref{cut}\,(d) the 3-momentum and invariant mass cuts are held fixed and the angular cut is varied. From  Figs.\,\ref{cut}\,(b)-(d) it is clear that the cuts have more effect on the decay width for the low Higgs boson mass. Fig.\,\ref{cut}\,(d) illustrates that the angular cuts only start to show any noticeable effect on the decay width $\Gamma$ for $\theta_{\rm cut} \gtrsim \pi/6$.

\section{Summary and Conclusions}

Compared to the one loop decays $H\to\gamma\gamma$ and $H\to\gamma Z$, the decay $H\to\gamma\gamma Z$ is highly suppressed in the standard model. In one sense this is unfortunate because the physics of this process has some interesting features. Among these are the absence of $W$ boson contributions in any of the loops. The amplitudes are dominated by top quark loops and therefore sensitive to top-$Z$ couplings.  Additionally, there is the presence of an anomalous vertex in the $s$-channel $Z$ exchange diagram, which might be studied were it not for the constraints of Yang's theorem \cite{yang}. Because Yang's theorem forbids the physical decay $Z\to\gamma\gamma$, the $s$-channel pole  cancels in the mass region we consider, $115\,\mathrm{GeV}\leq m_H\leq 160\,\mathrm{GeV}$. This, together with the anomaly cancellation, serves to further reduce the decay width. 

Perhaps the most striking feature of this process is the simplicity of the decay amplitude helicity structure. Although not apparent at the outset, the calculation shows that there are only two helicity amplitudes of any importance, ${\cal A}_{++0}=-{\cal A}_{--0}$. As a consequence, the photons in $H\to\gamma\gamma Z$ always have the same helicity and the $Z$ is always longitudinally polarized. 

In conclusion, we have found that $\Gamma(H\to\gamma\gamma Z)$ is very small compared to viable one loop discovery modes of the Higgs boson and that the suppression is greater than simple phase space and coupling constant accounting might suggest. It goes without saying that the detection of the decay mode $H\to\gamma\gamma Z$ will be extremely difficult if it occurs at the standard model level and that any signal in this channel is very likely evidence of new physics.

\begin{acknowledgements}
 One of us (A.A.) wishes to thank the Department of Physics and Astronomy at Michigan State University for its hospitality and computer resources. This work was supported in part by the National Science Foundation under Grant No. PHY-02744789.
\end{acknowledgements}

\end{document}